\documentstyle[12pt,epsfig]{article}

\begin{document}

\begin{center}\Large
{\bf Supercritical Disk Accretion with Mass Loss}
\end{center}
\vspace{15mm}
\begin{center} Galina V. Lipunova \\
Sternberg astronomical Institute,\\
Universitetskii pr. 13, Moscow, 119899, Russia\\
~~ \\
{\it galja@sai.msu.su}
\end{center}
\vskip 2cm
\section*{Abstract}
We consider supercritical accretion onto a compact object when
the accretion rate exceeds its value derived from the Eddington limit:
$\dot M_{Edd}=L_{Edd}/c^2$. We use the scenario that was proposed by
Shakura and
Sunyaev (1973) for the supercritical regime, according to which matter 
flows out of the gas-disk surface under radiation pressure. An analytic
solution of the accretion-disk structure is obtained for this scenario by
using classical equations. An attempt is made to
construct a model for nonconservative supercritical accretion with  
advection. To this end, we develop a numerical scheme for calculating the
structure of an advective disk with mass outflow. The goal of solving the
problem of a supercritical disk with mass loss is to explain accreting   
sources with outflowing envelopes, for example, SS~433 and galactic nuclei.
The structure and observational manifestations of the envelope which forms 
around a supercritical disk are considered by taking into account
bremsstrahlung absorption and Thomson scattering of photons.

\newpage
\section*{Introduction}
The accretion-disk theory is used in astrophysics to explain a wide range of
X-ray sources, their energetics, spectra, and variability. By now a wide
variety of models have been proposed to account for the observed
characteristics of quasars, X-ray pulsars, and cataclysmic variables.

The disk-accretion studies by Gorbatskii (1964), Lynden-Bell (1969), Shakura
(1972), Pringle and Rees (1972), and Shakura and Sunyaev (1973) yielded the 
first mathematical theory of accretion disks. The general pattern of
steady-state accretion in disks is as follows: the gravitational energy of
the gas that rotates around the central star transforms into the thermal and
kinetic energy of orbital motion through friction. For the effective
emission of heat, the gas flows cool down and produce a disk structure. The
disk luminosity is a function of the central object's mass, the rate of
accretion onto it (i.e., the rate of mass inflow to the outer boundary of
the accretion disk from the interstellar medium or from a neighboring star),
and the specific mechanism of angular-momentum transfer in the disk.

When the accretion rate is not too high, the total accretion-disk luminosity
is directly proportional to the accretion rate. Under such quiet conditions,
the global disk structure can be successfully described by the classical
solution of Shakura and Sunyaev (1973).

However, it is well known from the scenario for the evolution of binaries
(Tutukov and Yungelson 1973; Van den Heuvel 1994) that there are
evolutionary stages of violent mass transfer between the component stars in
binary systems. At $\dot M$ greater than the critical value
\begin{equation} 
\dot M_{cr}=L_{Edd}/{\vartheta c^2} \simeq 2\times 
10^{-9} ~ \frac{m}{\vartheta} ~M_{\odot}/\mbox{yr}~,
\label{Mcrit}
\end{equation}
where $m=M/M_{\odot}$ is the stellar mass (in solar masses), and
$\vartheta$ is the
accretion efficiency in a thin disk, the disk luminosity becomes so large  
that the radiation pressure begins to hamper the infall of matter to the   
center. This accretion regime is called supercritical or super-Eddington. In
a spherically symmetric case, the luminosity reaches the Eddington limit,   
when the gravitational attraction by the central star equals the radiation  
pressure on a proton:
\begin{equation}
{L_{Edd}\over {4\pi r^2}}~{\sigma\strut_T\over c} = {m_p M G\over {r^2}}~,
\qquad L_{Edd} \approx 1.25 \times 10^{38} m~ \mbox{erg}/\mbox{s}~,
\label{Eddington}
\end{equation}
where  $\sigma\strut_T=6.65\times 10^{-25}\mbox{cm}^2$ is the cross section
for Thomson scattering by free electrons.

In reality, the accretion rate clearly exceeds the critical value
(\ref{Mcrit}) in
several cases. For example, in SS~433, matter from the supergiant flows onto
the compact object (a neutron star or a black hole) at a rate of
$\sim 10^{-4} M_{\odot}$~yr$^{-1}$ (Cherepashchuk 1981). High accretion rates may occur in galactic
nuclei during intense infall of stellar matter onto supermassive black holes
(see, e.g., Dopita 1997; Szuszkiewicz et al. 1996).

Shakura and Sunyaev (1973) proposed a self-regulation mechanism for
supercritical accretion. As soon as the luminosity of the disk exceeds the
Eddington limit, matter flows out of its surface, causing the rate of
accretion onto the central object to decrease. The amount of gravitational
energy which can be liberated in the disk thus decreases, and the disk
luminosity is limited by a value that does not greatly exceed the Eddington
limit.

The outflow from the disk surface takes place inside the radius at which the
disk thickness is equal in order of magnitude to the disk radius. The disk  
structure outside can be described by the classical Shakura--Sunyaev
solution. This radius $R_s$ is called the spherization radius. As was 
established by Shakura and Sunyaev (1973), if the accretion rate in the disk
decreases as
\begin{equation}
\dot M (r) = \frac{r}{R_s} \dot M_o~,
\label{flowSHS}
\end{equation}
then the total disk luminosity does not greatly exceed the Eddington limit:
$L_{tot} \sim L_{Edd}\times (1+ln ({\dot M_o / {\dot M{_{cr}}}}))$~.
In this case, the total mass loss rate in the wind is
$\sim \dot M_o (1-\dot M_{cr}/\dot M_o )~$,
where $\dot M_o$  is the initial accretion rate at the outer disk radius.

The hypothesis of mass outflow from the disk is confirmed, for example, by
photometric and spectroscopic observations of SS~433, which suggest the   
presence of an envelope of matter flowing out of the disk (e.g., Murdin et
al. 1980). In some galactic nuclei, observations reveal outflow rates
reaching $\sim 1 M_\odot/$yr  (Heckman et al. (1981).

The authors who consider the winds from accretion disks, for example, in
cataclysmic variables, propose the mechanism of wind generation in spectral
lines. The rate of mass loss by such disks is
$\sim 10^{-11\div12}~M_{\odot}/$yr at a  
flow rate of $\sim 10^{-8}~M_{\odot}/$yr
(see, e.g., Suleimanov 1995; Pereyra et al. 1997). By contrast,
for supercritical mass outflow, a large optical depth in the disk
and the wind gives rise to a blackbody
spectrum and causes this mechanism to decrease in importance. Meier (1979)
considered the wind generated by a supercritical disk at a certain radius 
close to the spherization radius (for the Shakura--Sunyaev disk model) and
computed the vertical structure of the disk and the wind. Here, we consider
the radial structure of an outflowing disk with height-averaged physical   
parameters. For SS~433, our outflow rate is comparable in order of magnitude
to the value of Meier (1982): $3\div 9\times 10^{-6}~M_\odot/~$yr.

\section*{Equations for disk accretion with mass loss}

The structure of accretion disks at subcritical accretion rates is described
by standard equations of disk accretion (Shakura and Sunyaev 1973). In this 
case, the disk geometry can be roughly described by a thin disk, and
appropriate simplifying assumptions, for example, replacement of the
physical parameters in the disk by their height-averaged values, can be
imposed on the equations. To describe a nonconservative disk --- a disk with
mass outflow from its surface, we use the equations from the standard model,
which we modified to fit our view of mass outflow from the disk.

The equation of radial angular-momentum transfer along the disk --- the
\hbox{$\varphi$ component} of the Euler equation integrated over the coordinate  
perpendicular to the disk plane, is
\begin{equation}
{d\over dr}\{\dot{M}(r)\omega r^2\}= -2\pi{d\over dr}\{W_{r\varphi}r^2\}
                                  + {d{\dot{M}(r)}\over dr}\omega r^2~,
\label{MomTransf1}
\end{equation}
Here, $\dot{M}(r)$ is the accretion rate, $r$  is the radius,
$\omega$ is the Keplerian angular velocity, and $W_{r\varphi}$ is
the integrated, height--averaged  tangential component of the
stress tensor $w_{r\varphi}$ in the accretion disk:  
\begin{equation}
 W_{r\varphi}=2H w_{r\varphi}\qquad
 w_{r\varphi}=-\eta r \frac{\partial \omega}{\partial r}~,
\label{tensor}
\end{equation}
where $H$ is the disk half-thickness, and $\eta$ is the dynamical viscosity.
The change in angular momentum of the accreted gas is the sum of the
angular-momentum transfer by viscosity (the first term on the right-hand
side of (\ref{MomTransf1})) and the transfer of
angular momentum carried away by the
outflowing matter (second term). This equation differs from the standard
equation of angular-momentum transfer (see Eq. (\ref{MomTransf})),
in which  $\dot{M}=$~const,
precisely by the presence of the second term on the right-hand side.

The turbulent viscosity can be described by the dynamical coefficient,
$\eta\approx\rho < v_t l_t >$~,
where $v_t$ and $l_t$ are the characteristic velocity and
length of turbulent mixing, respectively. Using Prandtl's semiempirical  
relation
$$v_t =l_t r\frac{d\omega}{dr}~,$$
we find from (\ref{tensor}) that
$$w_{r\varphi}\simeq -\rho v_t^2=- m_t^2~\rho a^2~,$$
where $m_t$  is the Mach number, $m_t^2=v_t^2/a^2~,$,
and $a$ is the speed of sound.
Shakura (1972) introduced a turbulence parameter $\alpha$, which relates the
tangential component of the stress tensor to the pressure in the disk
$P=P_{gas}+P_{rad}$ (the sum of the gas and radiation pressures):
\begin{equation}
  w_{r\varphi}= - \alpha P~.
\label{turbul}
\end{equation}
The disk models that assume turbulence to be the source of viscosity and
that use relation (\ref{turbul}) are called $\alpha$--disks.

The equation of energy balance: the mechanical energy due to viscosity
produced in the disk at radius $r$  is completely emitted at the same radius,
\begin{equation}
  Q^+(r) = Q_{rad}(r)~,
\label{balans}
\end{equation}
where $Q_{rad}(r)$ is the radiative flux from one of the two accretion-disk
surfaces, and $Q^+(r)$~($Q^+(r)$~(${ {\mbox{erg}}/ {\mbox{cm}}^2
{\mbox{s}}}$) is half the heat released in a disk annulus of width $dr$.
\begin{equation}
Q^+(r) = -{3\over 4} \omega W_{r\varphi}~, \qquad
   Q_{rad}= {1\over 3} {c\over k_T \rho H} \varepsilon_{rad}~,
\label{fluxes}
\end{equation}
where $k\strut_T=\sigma\strut_T/m_p= 0.4~ \mbox{cm}^2/$g
is the specific Thomson scattering cross
section ($m_p=1.67\times 10^{-24}$~g is the proton mass),
$\varepsilon_{rad}$ is the mean radiative energy density,
and $\rho$ is the mass density. Equation (\ref{balans}) requires
special stipulations in the case of a  high accretion rate:
it turns out that the heat exchange between a given portion of the disk and
its adjacent portions must also be taken into account, as discussed below. 
Note that the ``local" equation of energy balance (\ref{balans})
allows an analytic treatment of the problem with outflow.

The expression for $Q_{rad}$ in (\ref{fluxes})
is the solution of the transfer equation in
the diffusion approximation, which is valid for optically thick disks. The 
optical depth $\tau$ of such a disk, whose geometrical thickness can be    
determined from the condition that the vertical component of the gravity is
equal to the pressure (hydrostatic equilibrium),
\begin{equation}
{1\over{\rho}}{dP\over{dh}}\sim {P\over{\rho H}}= {GMH\over r^3}~,
\label{hydro}
\end{equation}
is much larger than unity at supercritical accretion rates. Here, we
disregard the general-relativity effects. Novikov and Thorne (1973) 
considered thin accretion disks around Schwarzschild black holes in terms of
the general--relativity theory.

The equation of state
\begin{equation}
P=\rho a^2={\varepsilon_{rad}\over 3}
\label{press}
\end{equation}
holds in the radiation-dominated regions of thin disks. In the supercritical
regime, the radiation pressure in the disk is much higher than the gas
pressure.

The spherization radius $R_s$ at which mass outflow sets in can be
approximately determined from the condition that the luminosity reaches the
Eddington limit (Lipunov 1992):
\begin{equation}
{1\over 2} {GM \dot M_o\over{R_{s}}}\approx L_{Edd}~,
~~~ \mbox{then we have}~~~
R_s\approx R_o \frac{\dot M_o}{\dot M_{cr}}
~,
\label{rsfer}
\end{equation}
where $R_o$ is the radius of the inner disk boundary.

The  matter flowing out of the disk surface produces an expanding
envelope. The problem of mass outflow from a supercritical accretion
disk is similar, except for the geometry, to the problem of outflow of
strong winds from WR stars and giants. The wind structure is described
by the continuity, momentum, and energy equations. We do not set the  
goal of solving this complex problem and use an upper limit to
determine the outflow rate. For the matter from the disk to be moved to
infinity, the work
 $ d \dot M \left(\frac{GM}{r}-\frac{V_{\rm Kepl}}{2}\right) $
must be done in a second. Clearly, an upper limit on $ d\dot M$
can be placed by assuming that this power is equal to $ 2~Q_{rad} 2 \pi r dr$,
the luminosity of the disk annulus from the two  
surfaces. We thus obtain the following estimate, which is the so-called
``energy-wind" relation:
\begin{equation}
{1\over 2} {d \dot M\over{2\pi r dr  }} {(v_{\infty}^2-v_{Kepl}^2)\over{2}}
\equiv
{{\omega^2 r}\over 8\pi} {{d\dot{M}(r)}\over{dr}} = Q_{rad}~.
\label{flow}
\end{equation}

For the mass outflow from the disk, the accretion flux decreases with radius
and is the sought-for function which is given by
\begin{equation}
  \dot M (r) = 4 \pi  r H\rho v_r~,
\label{defin}
\end{equation}
where $v_r$ is the radial velocity in the disk.

\section*{Conservative solution}

The solution to the set of equations (\ref{MomTransf1})
(without the second term on the right-hand side)--(\ref{press})
is a classical solution for a thin accretion disk
without outflow, which is called the standard model. In this case, the  
accretion efficiency is $\theta=1/12$. The equations can describe the disk
structure in different regions, depending on the form of equation of state,
i.e., on the relative contribution of the radiation pressure and the gas   
pressure. The solution to the equation of angular-momentum transfer,
\begin{equation}
{d\over dr}\{\dot{M}\omega r^2\}= -2\pi{d\over dr}\{W_{r\varphi}r^2\}~,
\label{MomTransf}
\end{equation}
for  $\dot M=$~const and the Keplerian law of motion
$\omega = \sqrt{GM/r^3}$ is the following (Shakura and Sunyaev 1973)
\begin{equation}
 W_{r\varphi}=\frac{\omega \dot M}{2\pi}
\left(1-\sqrt{\frac{R_{in}}{r}}\right)+
W_{r\varphi}^{in}\left( \frac{R_{in}}{r} \right)^2~,
\label{VRFI}
\end{equation}
where $R_{in}$ is the inner radius, and
$W_{r\varphi}(R_{in})\equiv W_{r\varphi}^{in}$~. In the
standard solution at the inner disk boundary $R_{in}=R_o$, one assumes that  
$W_{r\varphi}^{in}=0$.

The total disk luminosity in this solution is proportional to the accretion
rate,
\begin{equation}
L_{tot} = {1\over 2}{\dot M GM\over{ R_o}} =
L_{Edd}~ {\dot M\over{\dot M_{cr}}}~, \quad       \dot M < \dot
   M_{cr}~.
\label{Ltotclas}
\end{equation}

If the central star
is a Schwarzschild black hole, then  $R_o=3 R_g $, the radius of the
marginally stable orbit; if it is a neutron star, then the inner disk
boundary lies either on the neutron-star surface\footnote{For a
neutron star with a soft equation of state, the stellar radius may turn
out to be smaller than the radius of the marginally stable orbit.} or 
at the Alfven radius in the case of a strong magnetic field near the   
neutron star.

The disk thickness in the regions under consideration changes with radius
only slightly, as we show below. The condition for radiation domination can
be written as the equality of the forces of attraction and radiation
pressure on the disk surface,
\begin{equation}
\frac{m_p G M H}{r^3}\approx \frac{Q_{rad} ~\sigma\strut_T}{c}~.
\label{lokEdd}
\end{equation}
Using (\ref{balans}), (\ref{fluxes}), and (\ref{VRFI}),
we obtain $Q_{rad}(r)$ and, after simple algebraic transformations, derive
\begin{equation}
H= R_g \frac{\dot M}{\dot M_{cr}} \frac{3}{4\vartheta}
		\left(1-\sqrt{\frac{R_{o}}{r}}\right)\approx 
   R_g \frac{\dot M}{\dot M_{cr}} \frac{3}{4\vartheta} ~(\mbox{при}~r>>R_o) ~.
\label{tolschina}
\end{equation}

\section*{Solution in a nonconservative model}
The set of equations (\ref{MomTransf1})--(\ref{defin})
describes the disk structure within the
spherization radius, where mass outflow proceeds. Its solution can be found
analytically. From Eqs. (\ref{MomTransf1}), (\ref{balans}), (\ref{fluxes}),
and (\ref{flow}), we derive $\dot M(r)$ and $W_{r\varphi}~$: the accretion
rate in the disk, starting from the spherization
radius $R_s$ at which outflow sets in, falls off to the center as
\begin{equation}
\dot{M}(r)=\dot{M_o}(r) \left({R_s\over r}\right)^{3/2}
          { {1+{3\over 2}\left(\frac{r}{R_o}\right)^{5/2}}\over
            {1+{3\over 2}\left(\frac{R_s}{R_o}\right)^{5/2}} }~ ,
             \quad r \leq R_s \, .
\label{Mdot}  
\end{equation}
This law has the asymptotic limit
$$ 
\dot{M}(r)\approx  \dot{M_o}(r){r\over R_s} \quad 
\mbox{\hskip 0.5cm при \hskip 0.5cm} R_o<<r\leq R_s~,
$$
i.e., the derived solution matches Shakura--Sunyaev's solution
(Eq.~(\ref{flowSHS})). The analytic expression for the height-averaged
component of the stress tensor in the disk is
\begin{equation}
W_{r\varphi}={\hskip 0.1cm}-\frac{\dot{M}(r)\omega}{4\pi}
~\frac{1-\left(\frac{r}{R_o}\right)^{5/2}}
{1+\frac{3}{2}\left(\frac{r}{R_o}\right)^{5/2}}
~.
\label{WRFI}  
\end{equation}
The boundary conditions at the inner radius $R_o$ are $W_{r\varphi}=0$
and $Q_{rad}=0$. The absence of stress at the inner disk edge is undoubtedly a
simplifying assumption, which ignores the hydrodynamic pattern of the flow.

The full solution is obtained by joining the solution with outflow and the
solution without outflow (outside the spherization radius), which is the  
standard Shakura--Sunyaev solution for the appropriate boundary conditions
at the spherization radius $R_s$ (Eq.~(\ref{VRFI})).

The spherization radius is a parameter of the problem. It can be numerically
adjusted in such a way that the disk luminosity outside this radius is of   
the order of $L_{Edd}$~:
\begin{equation}
 R_s~:~~~ L(r> R_s,~ R_s) =L_{Edd}~.
\label{defRs}
\end{equation}
Note that the refined spherization radius gives a different total rate of
mass outflow from the disk. Bisnovatyi-Kogan and Blinnikov (1977) studied
the dynamics of particle motion near a transcritical disk and found that the
process of particle escape to infinity under radiation pressure in the
gravitational field of a central object or disk spherization is established
at a total disk luminosity in the range $0.6$ to $ 1 L_{Edd}$.
It was found that up
to 80\% of the matter could be carried away from the disk for $ 0.6 L_{Edd}$
However, since $R_s$ is currently rather uncertain, we use condition
(\ref{defRs}) in the numerical procedure.

In Fig.~\ref{struktura}, the solid lines represent some
of the disk characteristics as a
function of radius for $\dot m\equiv\dot M/ \dot M_{cr}=1000$;
the change in accretion rate, which is described by Eq. (\ref{Mdot}),
is shown in the lower left panel; the     
derived spherization radius is
$R_s=1.62 ~R_o~\dot M_o / {\dot M_{cr}}$ (cf. Eq. (\ref{rsfer})).

The upper left panel shows a plot of the relative half-thickness $H/r$ in the
disk against radius, which is constant (solid line) for $r < R_s$ and is the 
same for any $\dot M_o> \dot M_{cr}$. This can be easily
verified by substituting the accretion rate $\dot M(r)$, which decreases
under the effect of the energy-wind   
mechanism (\ref{Mdot}), into the expression for the disk thickness in the
radiation-dominated region (\ref{tolschina}).

\section*{Advective disks with mass loss} 
\subsection*{Formulation of the Problem and Results}
\begin{figure}[p]
\centerline{\epsfig{file=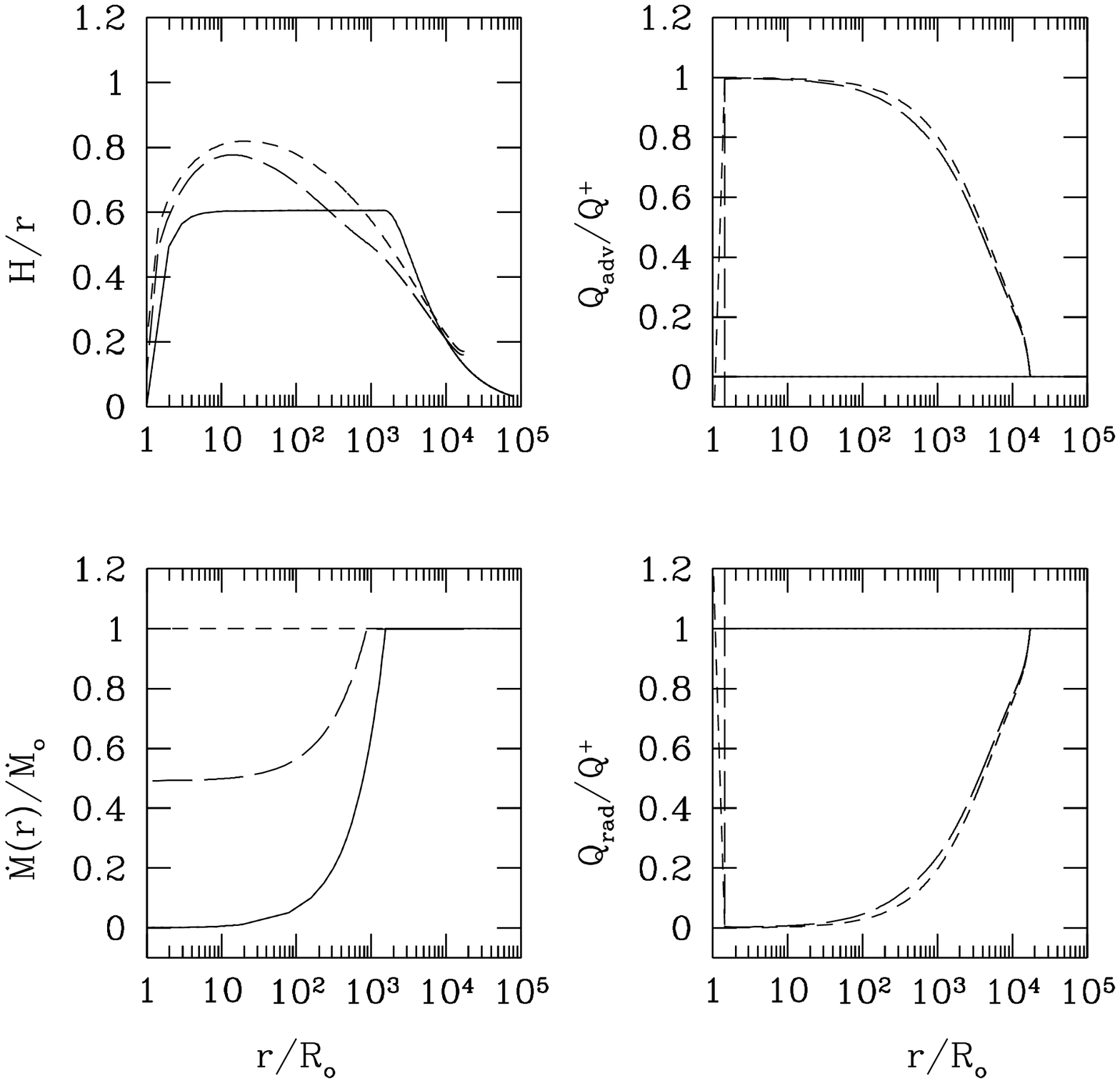,width=6in,height=5in}}
{\caption{
Relative thickness, accretion rate, fractions of emitted heat
and advection versus radius. The
solid line, the line with short dashes, and the line with long dashes
represent an analytic solution with outflow, and advective conservative
disk, and an advective nonconservative disk, respectively. The distance
is normalized to the inner disk radius; $\dot m=1000$.}
\label{struktura}}
\end{figure}

At supercritical accretion rates, the disks are always thick. The assumption
that $H/r <<1$ is invalid for them. This can be illustrated by an analytical
solution for a nonconservative disk: we see from Fig.~\ref{struktura}
(upper left panel,
solid line) that the relative half-thickness reaches $\sim 0.6$.

\begin{figure}[p]
\centerline{\epsfig{file=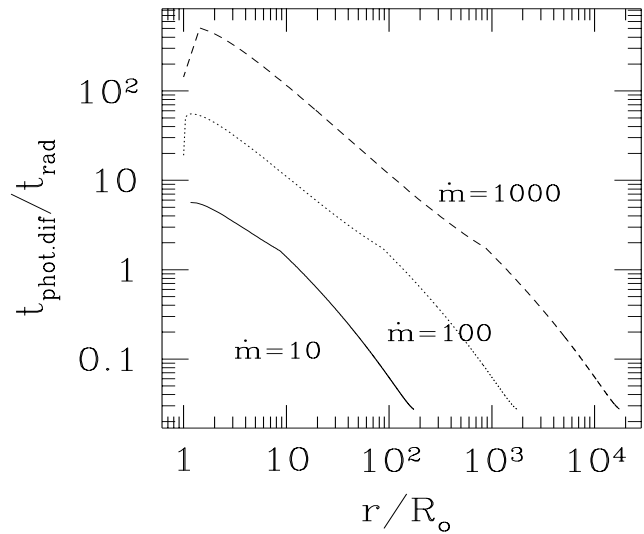,width=3.5in,height=3in}}
{\caption{
Ratio of the time of photon diffusion to the surface to the
characteristic time of radial matter motion in the disk versus radius. The
numerical calculations were performed for a nonconservative disk with
advection; $\dot m=$10, 100, 1000 and $\alpha=0.5$.}
\label{td_tr}}
\end{figure}

In a thick disk, the emission from the surface is not an efficient cooling
mechanism. In the characteristic time of inward matter motion along the   
radius, the heat cannot radiate away and is transported with the matter,  
i.e., the time it takes for a photon to reach the disk photosphere,
$ t_{\mbox{dif}}=3 H \tau/c$ (here, $\tau=H\bar n \sigma\strut_T>>1$
is the optical depth for scattering across the   disk, and $\bar n$
is the height-averaged number density of the disk matter), is 
much longer than the characteristic time of radial particle motion $r/v_r$, 
where $v_r\sim \alpha \left({H\over r}\right)^2 \omega r$ is the radial
velocity with which the matter moves over the disk (see Fig.~\ref{td_tr}).

This transport of heat along the disk with matter is called advection. If
advection is taken into account, the equation of energy balance
(\ref{balans}) changes:
\begin{equation}
Q^+ = Q_{rad} + Q_{adv}~,\qquad Q_{adv}=\rho v_r H T {d S\over{d r}}~,
\label{balans1}
\end{equation}
where $T$ is the temperature, and $S$ is the specific entropy. Using standard
thermodynamic relations, the expression for advection can be rewritten as 
\begin{equation}
Q_{adv}=\rho v_r H ~\frac{d\left({\varepsilon \over \rho}\right) 
                      +  {P~d \left({1\over \rho}\right)}}{dr}  
~,
\label{adv}
\end{equation}
where $\varepsilon$ is the internal energy density of the gas, and $P$ is the
total gas pressure (averaged over $H$). Inside the disk, where the optical
depth is much larger than unity, the photons and particles are in
thermodynamic equilibrium, and $\varepsilon=AT^4$,
$A=7.58\times 10^{-15}\mbox{erg}/\mbox{cm}^3\mbox{К}^4$.
Since the radiation pressure dominates over the gas pressure
in supercritical disks, we have
\begin{equation}
P\approx P_{rad}={AT^4\over 3}~.
\label{davl}
\end{equation}

Thus, the set of equations (\ref{MomTransf1})--(\ref{turbul}),
(\ref{fluxes}), (\ref{hydro}), (\ref{flow}), (\ref{defin}),
and (\ref{balans1})--(\ref{davl})
describes an outflowing disk with advection. A numerical solution of the 
disk structure can be found by using a developed computer code, which has
two modifications: to compute the structure of an advective disk with and
without outflow. We tested the numerical scheme on the Shakura--Sunyaev  
standard-disk model.

The results of our calculations for the accretion rate
$\dot M_o/(12 L_{Edd}/ c^2)=1000$  are shown in Fig.~\ref{struktura}
(long dashes). The short dashes represent the results of our
calculations for a conservative advective disk: no mass loss by the disk,   
and the accretion rate is constant.

As we see, in the solutions with advection, it dominates in the inner disk
regions, i.e., only a small fraction of the heat is emitted:
$Q_{rad}/Q^+ << 1 $.
In this region, the disk cooling by the energy wind (Eq.~(\ref{flow})) is
inefficient. Nevertheless, it is clear that a considerable fraction of the
initial accretion rate can be lost by the disk (long dashes).

Our numerical calculations for various $\dot m$ in the range
$10^2$--$10^4$ yielded
\begin{equation}
L_{tot}/L_{Edd}\approx 0.6 +0.7  \ln \dot m~. 
\label{Lcalc}
\end{equation}

\begin{figure}[p]
\centerline{\epsfig{file=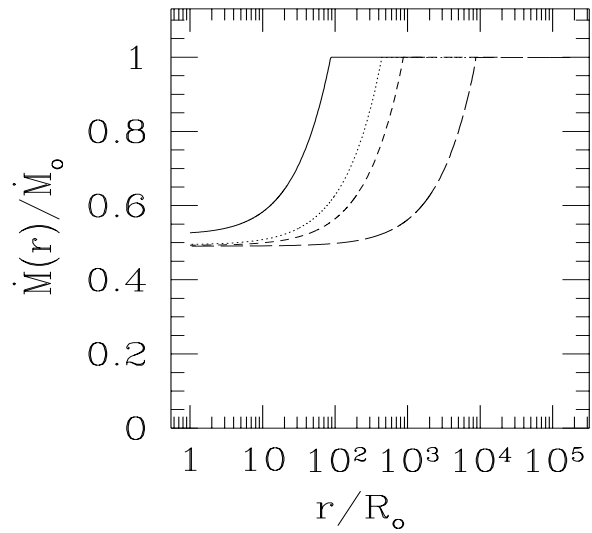,width=3.5in,height=3in}}
{\caption{
Change of the accretion rate in the disk due to mass outflow for
various initial accretion rates. The curves (from left to right) represent
the numerical calculations for an advective disk with
$\dot m=100,~500,~10^3,~10^4$.}
\label{dif_rates}}
\end{figure}

Figure~\ref{dif_rates} shows plots of the accretion rates
in the disk for various values  at the outer disk boundary.

Observations of SS~433  reveal an expanding envelope around the accretion
disk. The binary system SS~433 consists of an O or B star and a compact 
object (a black hole, $M\sim 10 M_\odot$,
or a neutron star, $M\sim 1 M_\odot$) (see, e.g.,
Van den Heuvel 1981). Mass accretion onto the compact object on the thermal
time scale of a normal star proceeds in the system at a rate of
$10^{-4}$--$10^{-3} M_{\odot}/$yr
(Cherepashchuk 1981; Van den Heuvel 1981). The mass
loss rate in the envelope is $10^{-5}$--$10^{-4} M_{\odot}/$yr
(Shklovskii 1981;
Dopita and Cherepashchuk 1981; Van den Heuvel 1981).

Thus, the observed outflow rate from the system, which is comparable to the
accretion rate, is obtained in the model of a supercritical nonconservative
disk.

\section*{Envelope around a supercritical disk}
\subsection*{Luminosity and Effective Temperature}
The envelope that is formed from the matter flowing out of a supercritical
disk can be optically thick for the emission from the inner disk regions. 
Its X-ray emission is reprocessed into a softer emission. Let us roughly  
estimate the effective temperature and emergent luminosity of the envelope
as a function of the model parameters and the empirically derived quantities
\begin{equation}
  A_1 = \frac{\dot M_{en}}{\dot M_o}~, \qquad
  A_2 = \frac{R_s}{\dot m R_o}~, \qquad   A_3 = \frac{R_o}{R_g}~. 
\label{koefs}
\end{equation}
The first two coefficients are determined when the disk structure is
calculated numerically; $\dot M_{en}$ is the mass loss rate in the envelope;  
$\dot M_o$ is the initial accretion rate at the outer disk boundary; $R_o$ is
the inner disk radius; $R_s$ is the spherization radius which is derived
from condition (\ref{defRs}); and the third coefficient is equal to 3, if the  
compact accreting object is a black hole, and is set approximately    
equal to 3 in the case of a neutron star:
$$
 A_1\simeq 0.5~, \qquad A_2=0.87~, \qquad A_3=3~.
$$
We assume that the mass outflow proceeds at a constant rate
\begin{equation}
\dot M_{en} = 4\pi R_s^2 m_p n_s v_{\infty}(R_s)~,
\label{neraz}
\end{equation}
where $n_s$ is the number density in the envelope at radius $R_s$,
and $v_{\infty}(R_s)=\sqrt{2GM/R_s}=0.62~ c~ \dot m^{-1/2}$
is the outflow rate.

\begin{figure}[p]
\centerline{\epsfig{file=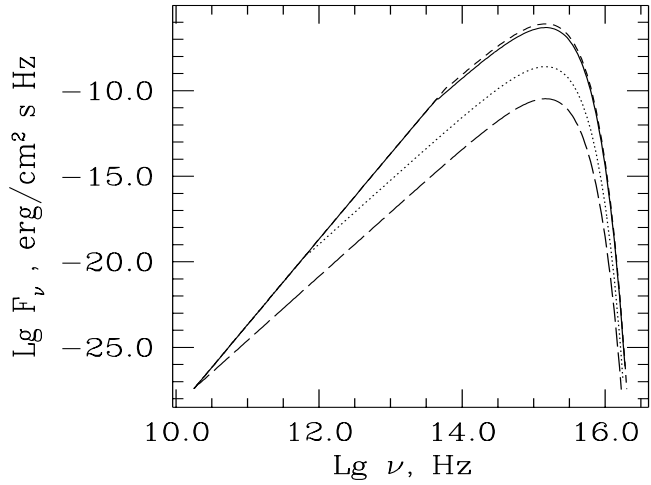,width=3.55in,height=3in}}
 \caption{
Spectrum of the envelope: a blackbody spectrum at low frequencies;
the emergent flux decreases with increasing frequency, when electron
scattering begins to dominate over bremsstrahlung absorption. The spectra
are given for $m=2$, $\dot m=5000$ and
$m=10$, $\dot m=1000$ (two upper close   
curves); $m=10^8$, $\dot m=10$ (middle curve); and
$m=10^8$, $\dot m=10^3$(lower
curve). The plasma temperature in the envelope is $T=3.5 \times 10^4$~K. The
temperature derived from Wien's displacement law is $\approx 2.5 \times 10^4 K$.}
\label{spectrum}
\end{figure}

\begin{figure}[p]
\centerline{\epsfig{file=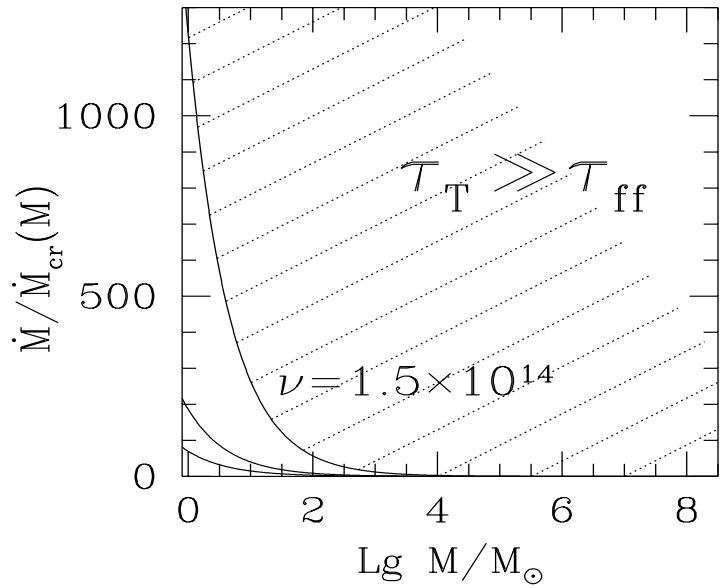,width=4.0in,height=3.5in}}
{ \caption{
Curves that represent the lower boundaries of the domains of ($m$,$\dot m$)
values in which the optical depth for scattering is larger than the  
optical depth for absorption at $\nu=10^{15}~$Hz (lower curve),
$\nu=4.5\times 10^{14}~$Hz (middle curve), and $\nu=1.5\times 10^{14}~$Hz
(upper curve).}
\label{tomson}}
\end{figure}

The emergent spectrum differs in shape from a blackbody spectrum, because
the scattering of emergent photons by free electrons plays a major role in
the opacity (Shakura and Sunyaev 1973; Sunyaev and Shakura 1974). The
visible envelope radius is $R\strut_T$, where the photon is scattered
for the last time. The radius $R\strut_T$ can be derived from the relation
\begin{equation}
\tau\strut_T(R\strut_T)\equiv 
\int\limits_{R\strut_T}^{\infty}\sigma\strut_T n dr
~=~\frac{\sigma\strut_T n_s R_s^2}{R\strut_T}=1~,
\label{R_o_def}
\end{equation}
by taking into account the fact that the particle number density in the
envelope at a constant outflow rate in it falls off as the inverse square of
the distance. From (\ref{koefs})--(\ref{R_o_def}) we obtain
\begin{equation}
R\strut_T=1.77\times 10^{6} A_1\sqrt{A_2 A_3} m \dot m^{3/2}~\mbox{cm}~.
\label{R_o}
\end{equation}
At frequencies at which scattering is larger than absorption, the emergent
spectrum from the last scattering surface is (Zeldovich and Shakura 1969)
\begin{equation}
F_{\nu}=\left(\frac{k_{ff}}{4\sigma\strut_T^2 R\strut_T}\right)^{1/4}\pi
B_{\nu}(T)~,
\label{outSpectr}
\end{equation}
where
$k_{ff}=3.69\times 10^8 T^{-1/2} \nu^{-3} g (1-e^{-h\nu/kT})~\mbox{cm}^5 $
is the bremsstrahlung absorption coefficient, $B_{\nu}(T)$ is the Planck
function, $g$ is the Gaunt factor, $T$ is the envelope
temperature, and $\nu$ is the frequency of the emergent emission. The shape 
of the spectrum determines the observed color temperature of the envelope.  
The temperature derived from Wien's displacement law (which is occasionally 
used instead of the color temperature) is mainly determined by the plasma   
temperature $Т$; $m$ and $\dot m$ actually affect only the flux and shape
while     
leaving the position of the peak in the spectrum unchanged (at those values 
at which the emergent spectrum is not a blackbody spectrum near the peak)   
(see (\ref{outSpectr}) and Fig.~\ref{spectrum}). The
frequency--integrated expression (\ref{outSpectr}) yields the  
flux $Q=54 R\strut_T^{-1/4}T^{25/8}\mbox{erg}/\mbox{s cm}^2$
(the Gaunt factor is set equal to 1) and is  
an analog of Stefan-Boltzmann's law (Sunyaev and Shakura 1974). The envelope
luminosity is then
\begin{equation}
L=4\pi R\strut_T^2 Q = 2.15 \times 10^{36}~ T_4^{25/8} R_{12}^{7/4}
~\mbox{erg}/\mbox{s}~,
\label{IntPotok}
\end{equation}
where $T_4=T/10^4$~K and $R_{12}=R\strut_T/10^{12}$~cm. The departures from
Stefan-Boltzmann's law become significant at
$T>2.3\times 10^3 R_{12}^{-2/7}$~K. Assuming that the spectrum has a
nonblackbody shape (i.e., when scattering dominates), if
$k_{\nu}<4\sigma^2\strut_T R\strut_T$, we find that this condition for
the emission at $\lambda=5500~$\AA ~takes the form
$\dot m > 186~m^{-2/3}$ (see also Fig.~\ref{tomson}). Using
(\ref{R_o}), we thus obtain
\begin{equation}
L=1.8 \times 10^{26} (A_1\sqrt{A_2 A_3})^{7/4} m^{7/4} \dot m^{21/8} 
T_4^{25/8}~\mbox{erg}/\mbox{s}~.
\label{IntPotok2}
\end{equation}

\subsection*{ SS~433}
Object no.~433 from the catalog of Stephenson and Sanduleak (1977) is famous
primarily for the two oppositely directed jets which move at a velocity of  
 $0.26 c$ (Fabian and Rees (1979). According to the current model, the jets
 are
generated at the center of an accretion disk around one of the components of
the close binary system, in which supercritical accretion takes place. Mass 
flow onto the compact object proceeds in the binary on the thermal time
scale of a normal star at a rate of $10^{-4}$--$10^{-3} M_{\odot}/$yr
(Cherepashchuk 1981; Van den Heuvel 1981). The mass loss rate in the
envelope is $10^{-5}$--$10^{-4} M_{\odot}/$yr (Shklovskii 1981; Dopita
and Cherepashchuk 1981; Van den Heuvel 1981). Assuming a lower limit
on the flow rate, we find from
(\ref{Mcrit}) that the following relation holds for SS~433:
\begin{equation}
m \dot m \approx 4.7\times 10^4 \vartheta~,
\label{mdotm}
\end{equation}
where $\dot m=\dot M_o/M_{cr}$, and $\vartheta$ is the accretion
efficiency in a thin disk (bears
no direct relation to the actual accretion efficiency, which is
$L_{tot}/(\dot M_o c^2)=1/\dot m (\vartheta L_{tot}/L_{Edd})$~).

For our estimates we consider two cases: (1)~$m= 10$, $\dot m=1000$;
and (2)~$m=1\div2$, $\dot m=10000$. However, a black hole and a neutron
star as the compact
object represent markedly different situations, because, in contrast to a  
neutron star, a black hole has an event horizon under which the matter and 
energy accreted in an advective disk can freely fall. From
(\ref{Lcalc}) we derive  
the total luminosities of outflowing advective accretion disks:
(1)~$L_{tot}\approx 5.4~L_{Edd, BH}\approx 7\times 10^{39}$~erg/s
(black hole) and
(2)~$L_{tot}\approx 7.0~L_{Edd, NS}\approx 9\times 10^{38}$~erg/s
(neutron star). For supercritical accretion onto a neutron star, the
bulk of the energy from the magnetospheric region can be carried away by  
neutrinos (Basko and Sunyaev 1976).

\begin{figure}[p]
\centerline{\hspace {3mm}\epsfig{file=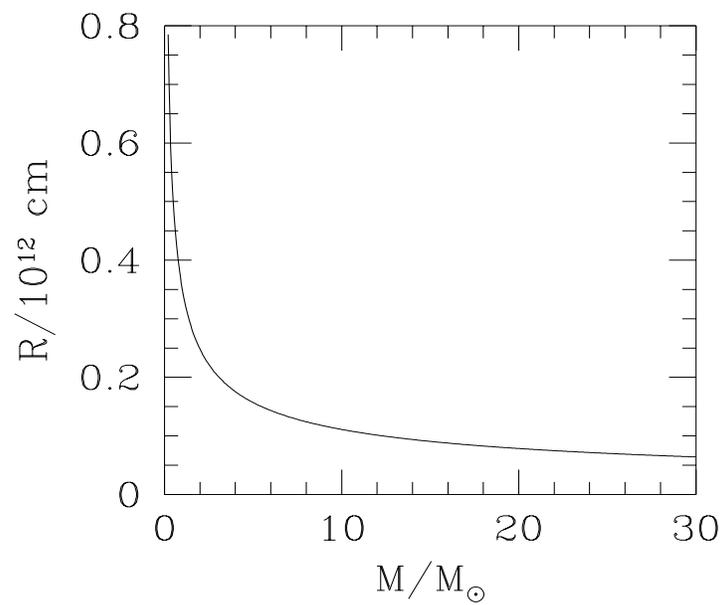,width=3.8in,height=3.5in}}
{\caption{
Visible envelope radius versus unknown mass of the compact
component in SS~433.}
\label{m-theta}}
\end{figure}

Using (\ref{mdotm}) and (\ref{R_o}), we can find a relation between
the uncertain mass of
the compact object in SS~433 and the radius of the visible envelope (see 
Fig.~\ref{m-theta}):
\begin{equation}
R_{12}\approx 14.6~\vartheta ^{3/2}  m^ {-1/2}~.
\label{m_theta}
\end{equation}
Various theoretical and observational estimates yield an expanding-envelope
radius of $\sim 10^{11}$--$10^{12}~$cm for SS~433 (see,
e.g., Fabrika 1984; Bochkarev  and Karitskaya 1985).

In the envelope of SS~433 for $m= 10$ and $\dot m=1000$ or even for $m=1\div2$
and $\dot m=10000$ , electron scattering dominates over absorption
(Fig.~\ref{tomson}). The
color temperature determined from the peak flux from SS~433 (Murdin et al.
1980) is $\sim 25 000$~K, which corresponds to the plasma temperature in the
envelope $T=3.5 \times 10^4$~K (Fig.~\ref{spectrum}).
From (\ref{IntPotok2}) we then derive $L\approx 3\times 10^{37}$~erg/s
for $m=10$ and $L\approx 1\times 10^{38}$~erg/s for $m=2$. This luminosity
estimate depends on the mass flow rate in the envelope. The formula
$v_{\infty}(R_s)=0.62~ c~ \dot m^{-1/2}$
yields $\sim 6000$~km/s for 10$M_\odot$. The observed Doppler widths of the 
``stationary" H$\alpha$ and He II lines are $\sim (1-2)\times 10^3$~km/s.

\section*{Conclusion}

Here, we considered supercritical accretion onto a compact object. As a
result, we obtained an analytic solution of the accretion-disk structure
using classical equations. We developed a numerical scheme for calculating
the accretion-disk structure with advection, i.e., with radial heat
transport by the accreted matter. We obtained an estimate of the visible
radius and luminosity of the envelope composed of outflowing matter as a
function of the rate of accretion onto a black hole by taking into account
bremsstrahlung absorption and Thomson scattering of photons.

The physical quantities in the equations we considered were averaged over
the height. Finding an exact solution is a difficult and interesting
problem, which requires considering two-dimensional hydrodynamic equations
and which may also be associated with the problems of the generation of jets
from sources with black holes.

\begin{figure}[p]
\centerline{\epsfig{file=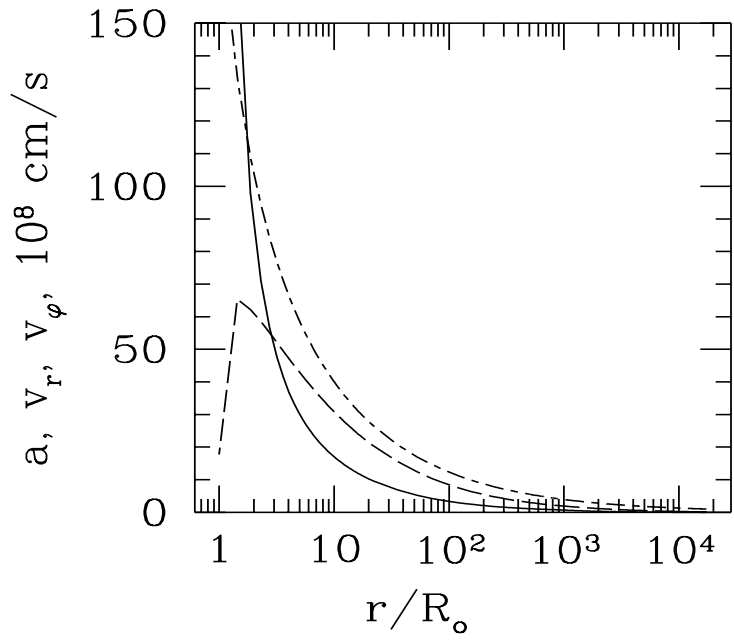,width=4.0in,height=3.5in}}
{ \caption{
Speed of sound $a$ (solid line), radial velocity in the disk $v_r$
(long dashes), and Keplerian velocity $v_\varphi$ (dot--dashed line) versus
radius. The numerical calculations were performed for a nonconservative disk
with advection for $m$ = 10, $\dot m$ = 1000, and $\alpha=0.5$.  }
\label{sonic}}  
\end{figure}

Here, we also assumed that the orbital particle velocity was described by
Kepler's law. In this case, we ignored the contribution of pressure to the
equation of motion (we took into account the pressure only when we
calculated the vertical disk structure, Eq.~(\ref{hydro})).
In distant regions, at $r>>R_o$, this is quite acceptable.
In Fig.~\ref{sonic}, the speed of sound, the radial
velocity over the disk, and the orbital Keplerian velocity
$\sqrt{GM/r}$ are plotted
against radius. Near the sonic point, i.e., at the distance at which the
speed of sound is equal to the radial velocity, $a=v_r$, the contribution
of pressure is large, and, as we see, the orbital and radial velocities are
of the same order of magnitude there, i.e., the particle orbits cease to be
Keplerian.

\begin{figure}[p]
\centerline{{\hspace {-1cm}}\epsfig{file=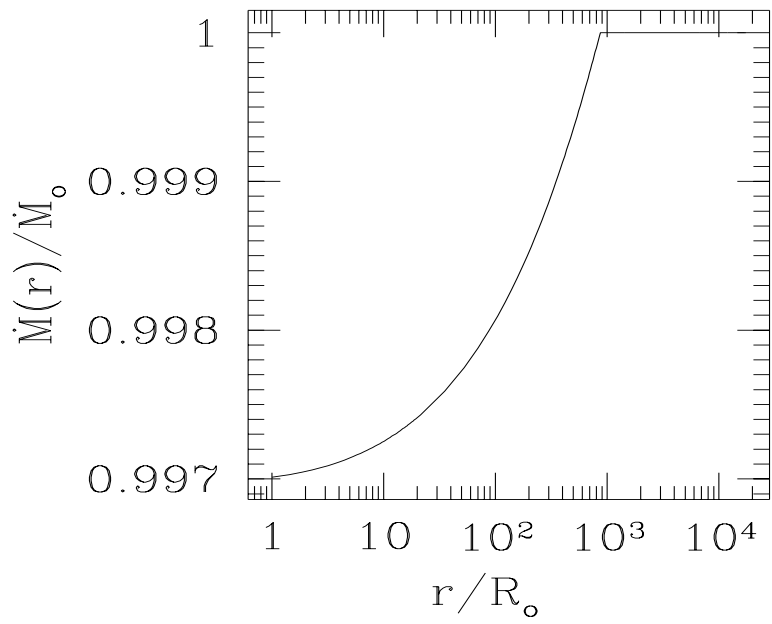,width=4.2in,height=3.5in}}
{\caption{
 Ratio of the accretion rate in the disk to the initial accretion
rate versus radius for mass outflow that follows the law
(\ref{imp_veter}). The
numerical calculations were performed for an advective disk with $m$ = 10 and
 $\dot m$ = 1000.}
\label{ImpFlow}} 
\end{figure}

The energy--wind equation differs from the relation that is used in the
standard stellar--wind theory,
\begin{equation}
m_p v\sim\frac{L}{c}~.
\label{imp_veter}
\end{equation}
If this law is used, then the efficiency of mass outflow from the disk
decreases by several order of magnitude. Figure~\ref{ImpFlow} shows
the results of our
numerical calculations for an advective disk with mass outflow that follows
the law (\ref{imp_veter}).
It turns out that using this law generally reduces the outflow
efficiency by more than two orders of magnitude and apparently cannot
explain the rates of mass outflow from SS~433, which are comparable to the
accretion rate in this system.

A distinctive feature of advective disks is that only a small fraction of
the liberated gravitational energy of the accreted matter is emitted, while
the heat is mainly transported in the disk. If the central star is a black 
hole, then matter together with energy fall into it without returning. By  
now the accretion-disk theory has progressed appreciably and has been
applied to observed sources. It turns out that advection can be of   
considerable importance in supercritical accretion and at low accretion
rates (see, e.g., Abramowicz et al. 1988, 1991; Narayan and Yi 1994; 
Narayan et al. 1997). In particular, in the opinion of these authors, the
model of a subcritical, optically thin, two-temperature advection-dominated
disk around a supermassive black hole can explain the observations of
low-luminosity galactic nuclei (Narayan et al. 1995; Lasota et al. 1996).

Observations of SS~433 show that the disk loses up to 90\% of the matter
flowing to its outer boundary, producing an outflowing envelope around the
compact star. Thus, the problem of calculating the structure of a
nonconservative accretion disk arises. In this paper, we considered an
approximate solution of this problem. We found numerical solutions to this
simplified problem (one-dimensional, etc.) with advection for a
wide range of accretion rates above
the critical value. We thus show that, in principle, substantial loss of
the accreted matter from the accretion disk is possible, and calculate 
its value.

\vskip 1cm
\par\noindent{\it ACKNOWLEDGMENTS}

I wish to thank N.I. Shakura, V.M. Lipunov, and M.E. Prokhorov for a helpful
discussion and the referees for useful critical remarks. I also wish to
thank V. Astakhov and N. Lipunova for help in preparing the paper.
This study was
supported by the International Program for Education in the Field of Exact
Sciences, the Russian Foundation for Basic Research (project no.
98-02-16801), and the Astronomy Science and Technology Program (1.4.2.3 and
1.4.4.1).


\end{document}